Working Draft

# Quantum Mechanical Foundations of Epistemology

Bruce Levinson

**Abstract**: Scholars of the history and philosophy of science have asked "What would decolonized science look like?" This paper develops an answer by interrogating Niels Bohr's assumption that observations need to be recorded and communicated using the language of classical physics. Bohr held this assumption even though he recognized the "fact that quantum phenomena cannot be analyzed on classical lines…." Classical analysis requires the objectification of whatever is being observed so that it can be objectively described. Inherent in this analysis is the intellectual creation of a transcendental verity, a "view from nowhere." In contrast, knowledge systems that predate European colonialism make use of the quantum nature of nature without constructing an objective model of it. This paper concludes that uncolonized knowledge systems differ from classical ones in that they are epistemically plural, accepting multiple true observations of the same phenomenon.

> *The fact that there were so few followers of Bohr is indicative of the havoc that a society set only on respect for the external can produce.*
>
> —Walter M. Elsasser[1]

## Introduction

What the hell are we talking about when we talk about quantum mechanics?[2] The question atomizes physicists. Adán Cabello recounts a journalist at a quantum physics conference who, after receiving multiple perplexing, sometimes metaphysical, usually mutually exclusive answers to his every question, asked the group how "do you plan to make progress if, after 90 years of quantum theory, you still don't know what it means? How can you possibly identify the physical principles of quantum theory or expand quantum theory into gravity if you don't agree on what quantum theory is about?"[3] The journalist could have asked the predicate question of how we got here in the first place. Why are there so many interpretations of the same equations?[4] Each interpretation, such as Copenhagen, is an exegesis of the mathematics that ties it to the physical reality we each experience. The factionalism is even odder given Richard Feynman's point that nature is quantum mechanical.[5] If nature itself is quantum mechanical and our model for understanding nature is quantum mechanical, why is there no consensus on the physical relation between the model and the reality that is being modelled? In the alternative, how could there be consensus? Mathematical models may be objective but human understanding is not. The growing diversity of quantum interpretations is inevitable and illustrates that individual understandings of theory—and of our shared reality—are inherently subjective. Proponents of the various interpretive schools posses the same knowledge of quantum theory yet hold very different beliefs about what that theory says about our physical world(s). Quantum physicists demonstrate that a wide range of contradictory beliefs are consistent with the same knowledge. In doing so, they demonstrate the distinction between quantum epistemologies and classical ones. Classical knowledge systems, such as conventional physics, are epistemically singular; they hold that there is a single truth about an observed phenomenon. There is little debate over the speed of light in a vacuum. The observation is independent of the observer. By contrast, the many interpretations of quantum theory suggest that epistemically plural knowledge systems support multiple "true" understandings of the same phenomenon. Pluralistic epistemologies do not create the distinction between knowledge and belief that is an axiom of modern thought. If Socrates and Gorgias had conversed after attending a quantum physics conference, their dialogue[6] may have gone something like this—

> Socrates: Then do you think that learning and believing, or true knowledge and belief, are the same thing, or different?
>
> Gorgias: In my opinion, Socrates, they are inseparable.
>
> Socrates: And your opinion is right, as you can prove in this way: if some one asked you—Does, Gorgias, the Schrödinger equation reflect true knowledge?—you would say, Yes, I imagine.
>
> Gorgias: I should.
>
> Socrates: But now, are there many conflicting understandings of this true knowledge?
>
> Gorgias: Surely yes!
>
> Socrates: So it is evident again that knowledge and belief cannot be separated.
>
> Gorgias: You are right.

Gorgias might agree, but Bohr does not. Niels Bohr argues that, from "the present point of view of physics...every description of natural processes must be based on ideas which have been introduced and defined by the classical theory"[7] which objectively distinguishes between knowledge and belief. He holds that language of classical physics this language is necessary for communicating scientific results no matter how "far quantum effects transcend the scope of classical physical analysis…." because

> the account of the experimental arrangement and the record of the observations must always be expressed in common language supplemented with the terminology of classical - physics. This is a simple logical demand, since the word "experiment" can in essence be used only in referring to a situation where we can tell others what we have done and what we have learned.[8]

Is Bohr correct? Must scientific communications about a quantum world rely on classical theory despite the "fact that quantum phenomena cannot be analyzed on classical lines..."?[9] This paper explores Bohr's assumption as it develops an answer to the question of what physics looks like when understandings of it are not constrained by classical Greek thought. It concludes that scientific systems unaffected by Greek thought make use of quantum phenomena, are epistemically plural—accepting multiple "true" observations of the same phenomenon, is it suggests an affirmative answer to Bohr's question of "whether its possible to present the principles of quantum theory in a way that their application appears free from contradiction."[10]

**A Story About Stories**

Objectivity is the foundation of modern science. It is counterintuitive to think that a scientific system could develop without researchers being able to communicate what each has done and what each has learned. Yet, even in an epistemically singular system, objective truths require elucidation and "every time a mathematician explains anything, how to find y, narrative is created"[11] and narrative is inherently personal—as the journalist at the physics conference discovered. The question is what sort of communication or encoding could convey individual observations in a form that would allow scientists to collaboratively create a scientific system, that is, a system for making and testing predictions about nature? One possibility is story. Humans make sense out of what our senses tell us via story. The language of mathematics is better suited to conveying abstract truths than the human experience of those truths. To explain experience, we need stories. This paper contends that oral traditions make use of stories and are able to "present the principles of quantum theory in a way that their application appears free from contradiction" as evidenced by their ability to accurately transmit observational data across generations.

For much of its existence, the Western academy has not recognized orally transmitted, knowledge. In written form, history goes back only about five thousand years. In its oral forms, history —the stories humanity tells about itself—records older human observations. Our oldest story may date back 100,000 years to when our common ancestors looked up at the east African sky and explained how it changed over time. Astronomers have recently uncovered previously unsuspected stellar movements in the Pleiades cluster that are explained in stories that the earliest humans created and kept alive as they migrated around the globe.[12] Observations made in the last 10,000 years or so that are subjectively described in Aboriginal stories from the Australian continent, have been objectively verified and dated through modern methods such as core sampling and radiometric analysis. These observations include experiences with environmental changes, encounters with now-extinct megafauna, and volcanic eruptions.[13] The term "stories" should be given a broad reading. For example, on the Australian continent the "nature of remembered Aboriginal history and geography is conveyed intergenerationally through stories, songs, dances and—more broadly— the 'law' of particular Aboriginal groups.…"[14]

The observations preserved in orally transmitted knowledge systems constitute a massive intergenerational transfer of information that scholars have largely ignored. Our failure, from a classical perspective, to recognize the epistemic processes involved in indigenous storytelling suggests that we may "have underestimated our ancient ancestors' cognitive abilities and, as a result, undervalued their capacity for innovation and misinterpreted some of their greatest achievements."[15] Since nature is quantum and oral traditions are able to communicate objectively verifiable observations over , this paper argues that oral-based knowledge systems make use of the principles explicated in quantum theory in a way that appears free from contradiction.

**An Observation About Observers**

Bohr tells us that we must take "great care" with our words since phrases in the physics literature "like 'disturbance of phenomena by observation'" use terms in ways that are not compatible with common understanding and are "apt to cause confusion."[16] An observer's observation can't "disturb" whatever is being observed because the observer, disruptive influence and all, is part of the observation, not independent of it. Efforts to ignore the role of the observer in the observation spawns the endless ontological bewilderment inherent in thinking that it is possible to stand outside of existence. The problem with concepts that are located in an "otherworldly Platonic realm of pristine forms uncontaminated by base matter"[17] is that they exclude the integral function of the observer in assigning meaning to what she is observing. The failure of medical science in Victorian-era England to recognize that observers attach meaning to data was observed by Susan Sontag in her discussion of how cancer has been linked to character.

> Snow was a surgeon in the Cancer Hospital in London, and most of the patients he saw were poor. A typical observation: 'Of 140 cases of breast-cancer, 103 gave an account of previous mental trouble, hard work, or other debilitating agency. Of 187 uterine ditto, 91 showed a similar history.' Doctors who saw patients who led more comfortable lives made other observations. The physician who treated Alexandre Dumas for cancer, G. von Schmitt, published a book on cancer in 1871 in which he listed 'deep and sedentary study and pursuits, the feverish and anxious agitation of public life, the cares of ambition, frequent paroxysms of rage, violent grief' as 'the principal causes' of the disease.[18]

The computational scientist Russell K. Standish explains that by "explicitly recognising a role for the observer of a system, an observer that attaches meaning to data about the system, contradictions that have plagued the concept of complexity can be resolved.[19] He notes that "[e]xplicitly acknowledging the role of the observer helps untangle other confused subject areas.... Quantum Mechanics can also be understood as a theory of observation. The success in explaining quantum mechanics, leads one to conjecture that all of physics may be reducible to properties of the observer."[20] Standish's conjecture is consistent with Bohr's late-in-life conclusion that physics "is to be regarded not so much as the study of something a priori given, but as the development of methods for ordering and surveying human experience."[21]

The irreducible role of the observer is emphasized by the mathematician Bruno de Finetti, who held an understanding of Bayesianism that is pure (or dogmatic) in its rejection of objective probability. The preface to de Finetti's Theory of Probability states that "PROBABILITY DOES NOT EXIST."[22] He further expounded "is it true that probability 'exists'? What could it be? I would say no, it does not exist."[23] De Finetti rejects the transcendental metaphysics that posits the "existence" of probability, an intellectual construct. To de Finetti, probability exists only as an individual's state of mind, without any corresponding physical meaning. In de Finetti's world, as in ours, bet-your-life (p=1) certainty does not equate to physical actuality—as obituaries routinely document. Complex thinking is the core of de Finetti's concept of probability. Instead of basing probability assessments on a given rule or regularity, they are based on the totality of an individual's experience and expectations. To de Finetti and other subjectivists "all coherent functions are admissible…."[24] De Finetti's focus on complexity and subjectivity brings us back to Standish's point that the role of the observer needs to be recognized in order for us to have a clear concept of complexity.

Science's dismissal of the role of the observer in attaching meaning to observations can be traced to the Greek *nomos-physis* dialectic with its distinction between "man" and "nature." However, although "we may now conceive of *physis* and *nomos* as antitheses, this was not always so. In the archaic phase of Greek thought, *nomos* is regarded as divinely willed *ipso facto* valid, normative order in harmony with *physis*….”[25] The change in Greek thought may have been influenced by Persia’s victory at Thermopylae a decade before Socrates was born. Dualist philosophy/theology was a major intellectual influence throughout the Persian empire but did not originate in Persia. Dualist metaphysics, which have become embedded in Western thought, can be traced back to “the Babylonian myth of the struggle between Merodach and the dragon, between the God of intelligence and the Tiamat monster of the chaos deep”[26] which remain in conflict through their progenies, science and art. In contrast with early Greek thought, the *nomos*-*physis* dialectic carries with it the implication that humanity is somehow set apart from the rest of the universe. Erwin Schrödinger recognized this metaphysical conflict as he wrestled with its implications. In doing so, he explained that he was retreading the same ground that Democritus had explored 2,400 years earlier in a dialogue discussing the existential crisis that occurs when Intellect reasons itself out of existence.

> (Intellect:) Sweet is by convention, and bitter by convention, hot by convention, cold by convention, colour by convention; in truth there are but atoms and the void.
> (The Senses:) Wretched mind, from us you are taking the evidence by which you would overthrow us? Your victory is your own fall.[27]

**Quantum Man, Quantum Knowledge**

Epistemically plural knowledge systems resolve Democritus’s existential difficulties by acknowledging that the observer’s senses are part of each unique observation. The problem with this metaphysical resolution is that it creates a practical problem that Bohr did not see a way around—how could an epistemically plural knowledge system work? How is it possible for a society’s knowledge system to be based on unique individual experiences, not objective data? Quantum mechanics, as a discipline, began in the early 20th century but the scientists who formalized explanations for their observations did not invent the quantum nature of the world and were not the first people to notice it. Nature being quantum mechanical means that we human beings are quantum mechanical too. It is now "indisputable...that the quantum dynamics that are undoubtedly taking place within living systems have been subject to 3.5 billion years of optimizing evolution" and it "is likely that, in that time, life has learned to manipulate quantum systems to its advantage in ways that we do not yet fully understand."[28] Since we are quantum, it is only "natural" that humanity's early knowledge systems are also quantum, that is to say, epistemically plural.

Humanity has produced many different knowledge systems—each of which comprehended and explicated the physical world with sufficient accuracy as to allow a society to rest on it, at least for a while. How do we reconcile the many different systems for creating and using knowledge of physical reality with a unitary concept of physical reality? One possibility is to recognize the shortcomings intrinsic in the classical understanding of reality. Schrödinger traced these shortcomings back to dualist Greek thought and opined that scientists had oversimplified their understanding of nature by omitting themselves from their models. He explained that when the scientist ignores the role of "his own personality, the subject of cognizance" in the study of nature, it "leaves gaps, enormous lacunae, leads to paradoxes and antinomies" whenever "one tries to find oneself in the picture, or to put oneself, one's own thinking and sensing mind, back into the picture."[29] Schrödinger further notes that

> This momentous step—cutting out oneself, stepping back into the position of an observer who has nothing to do with the whole performance—has received other names, making it appear quite harmless, natural, inevitable. It might be called just objectivation, looking upon the world as an object. The moment you do that, you have

> virtually ruled yourself out.... None the less, it is a definite trait. A definite feature of our way of understanding Nature—and it has consequences.[30]

Since the Greek atomists, scientists have presumed and continue to pursue an epistemically singular explanation for all phenomena. Modern science's founding metaphysical assumption unconsciously conveys the memory of a Babylonian myth that continues to have consequences. One of these consequences is the difficulty that physicists encounter in agreeing on how to comprehend the physical meaning of quantum theory. Another consequence of objectivation is that models of human behavior are less accurate than models which make predictions about the "natural" world. It seems that separating ourselves from the world we study results in a knowledge system in which we know less about ourselves than we know about everything else.

Yet, despite whatever flaws it may have, the classical model has come to dominate the world and not without reason, particularly genocide. For example, "[i]n Tasmania, colonists who seized Aboriginal children and dashed their brains out or lined Aborigines up as targets for musket practice likely considered their victims less than human. . . ."[31] In 1851, the Governor of California declared that a "war of extermination will continue to be waged ... until the Indian race becomes extinct."[32] During the 20th century, many countries carried out policies that were intended to eradicate indigenous peoples. These policies continue.[33] We can see from these—and many similar examples—that the global dominance of the epistemically-singular Western model of science is not based on it having a monopoly on understanding nature, that is, on making and testing predictions for navigating the world. Instead, it is the result of its success in eliminating the cultures that produce and sustain other knowledge systems.

Greek philosophy has been incorporated throughout modern thought so thoroughly that Niels Bohr considered the need for objectivity self-evident even though "quantum phenomena cannot be analyzed on classical lines….” In trying to answer the question of what physics looks like when it does not incorporate a *nomos-physis* dialectic, we compare  a classical understanding of knowledge embodying a distinction between the intellect and the senses and compare it with an epistemically plural *nomos* which integrates all aspects of a society’s knowledge into a unified system.

**The Intellect (Scientific Knowledge)**

Science, however it is understood in a given place and time, plays a crucial role in social structure. Since ancient Egypt, science has been inextricably entangled with law.[34] Every knowledge-generating community—every *nomos*—posses an understanding of physical reality and derives legal functions from it. Law may be expressed via equation, code, story, or ritual but a legal function encoding a scientific system for understanding the world is embedded in every *nomos*. The law teacher Robert M. Cover posits that “*each* ‘community of interpretation’ that has achieved ‘law’ has its own *nomos*”[35] and that this “*nomos* is as much ‘our world’ as is the physical universe of mass, energy, and momentum.”[36] In the absence of objectivation, the two are inseparable.

Most societies today are epistemically singular. Under the legal aspect of such systems, an infraction occurs at the time an actor’s actions violate the law. The infraction is an objective fact independent of whether it is observed. Epistemically plural knowledge systems take a more personal approach. Although he practised his profession in an epistemically singular society, Oliver Wendell Holmes brought an epistemically plural perspective to his job. Justice Holmes asks “[w]hat constitutes the law?”[37] His answer, “[t]he prophecies of what the courts will do in fact, and nothing more pretentious, are what I mean by the law,”[38] reveals an unusual understanding of the subject. Holmes is defining the law as a prophecy about the results of a contemplated experiment—a risky course of action. Will the prophesier be able to avoid punishment?

Holmes’s definition of law is de Finettian in that it is an individual’s subjective probability inferred at a given time, in a given situation with no objectively "true" value. Law is immanent and probabilistic under this definition, not external to the experimenter. Instead of being an ontic force, law exists only because bad people expect it to be enforced. Holmes rejects a transcendental concept of law in favor of one based on experience. He viewed human actions as bets on expected outcomes. In a 1929 letter to the jurist Frederick Pollock, he explained what he had learned from the philosopher Chauncey Wright—that "I must not say necessary about the universe….we don't know whether anything is necessary or not. So I describe myself as a *bet*tabilitarian. I believe that we can bet on the behavior of the universe in its contact with us."[39] Holmes kept up on scientific developments, as evidenced by his letter to Pollock, and was aware of "the new discoveries of quantum mechanics."[40] Holmes believed that there was a cause behind every phenomenon. Nonetheless, he recognized the novel ideas of scientists, such as Arthur Eddington, who were "suggesting...that there are phenomena for which no causes can be discovered and seemingly believing that they are outside the category of cause and effect." While Holmes was "far from believing" the physicists' speculations, he was "entirely ready to believe it on proof."[41]

Holmes's belief that law is determined on a personal level is a radical departure from epistemologically singular views, such as that law is what a statute says, or what a judge says, or what a ruler says, or what nature or a deity is perceived to be saying, all of which are external to the experimenter. Holmes's notion that truth emerges through a testing process and does not have an independent existence remains audacious. The subjective, emergent nature of Holmes's definition of law stands in stark contrast to Justitia's dictate that all persons are equal before *the law*. However, the Justice does warn us that "fashion is potent in science as well as elsewhere."[42]

One of the implications of Holmes's definition is that an action's legality is asynchronous with the action itself. Under Holmes's definition, an action is not unlawful until the court makes its decision which is after the action itself takes place. An actor's decision in the present, the now, is influenced by the "future," that is, a personal probabilistic set of an action's expected consequences, and by the "past," a probabilistic set of memories. Past and present can be understood as informational influences on an individual's decision-making process in the now rather than as concrete events that have or will occur. Holmes's approach to law is consistent with the physicist's perspective that "the distinction between past and future...is statistical only."[43] For physicists, it is "easier to think about the whole of spacetime as a single four-dimensional entity...in which temporal notions are a matter of perspective."[44] The gaps, lacunae, paradoxes, and antinomies that result from trying "to put oneself, one's own thinking and sensing mind, back into the picture" in an epistemically singular *nomos* include perplexity at our ability to "remember the past (but not the future)" and to "influence the future (but not the past)"[45] even though, in theory, we should be able to do both. Holmes shows us that we remember—obtain information from—the future via our expectations, an influence that is mediated through each individual's decision-making processes.

Our epistemically singular concept of time brings an additional difficulty, we don't know when now is. One of objectivation's consequences is that neuroscientists have questions about time that are outside the scope of the physicist's purview (and vice versa) but go to the core of human experience; questions such as what "is the subjective 'now', and what time window does it reflect—i.e, what is the 'width' of subjective now?"[46] Holmes implies that an individual's "now" occurs when an actor takes an action.

**The Senses (Poetic Knowledge)**

Poetic knowledge, knowledge based on the Senses, shares with scientific knowledge the goal of helping individuals make sense of and navigate the world. The difference is that scientists operate "patiently methodically, and inductively" while poets act "'intuitively, deductively, with large daring shortcuts and approximations.'"[47] The common goal raises the question of the two knowledge systems' relationship with each other. Cooperative? Competitive? Stone throwing?

From the epistemically singular scientific perspective, there can be the "problematic assumption that art can only be considered a form of knowledge if it conforms to scientific standards."[48] On the other hand, there is the concern that "an art or a poetics of knowledge can emerge that questions the actual construct of the sciences."[49] Science demonstrates its claim to knowledge through its accurate predictions about nature while poetic knowledge examines and subverts science's axioms. Poetic knowledge explores and tries to make sense of the peculiarities found in an epistemically singular *nomos*. It "questions the tight limitations of modern rationality by articulating — in contrast to objective, absolute, consistent scientific knowledge — knowledge that is equally ambivalent, incommensurable, and singular."[50]

The senses are a counterpoint to the intellect but can they constitute a knowledge system on their own? Music may provide an answer. Every *nomos* has its own culturally-specific musical traditions. These traditions share one commonality—music is epistemically plural, even in an epistemically singular *nomos*. There is no single truth about a musical experience. Does this mean that music is "quantum" or, in the alternative, that quantum theory is musical? Schrödinger pointed out that when multiple systems engage with each other and are then separated, the systems "can no longer be described in the same way as before"[51] which he called, not "*one* but rather *the* characteristic trait of quantum mechanics."[52] Although Schrödinger described this situation as "peculiar," it can be understood in a non-peculiar way. Consider two systems, the first composed of a violist, a cellist, a first violinist, and a second violinist, each with instrument. The second system consists of twelve children. When the systems engage with each other according to the law notated in Beethoven's String Quartet in E♭ major (Opus 127), the experience cannot be reproduced through knowledge of the separate systems and neither system emerges from the interaction unchanged.

**Epistemically Plural Knowledge**

Science excels at explaining how nature works but a poor job of explaining the human experience of it. Poetry is more useful for describing the experience of hunger than for alleviating it. Schrödinger warns us that by modelling the world as if we didn't exist, scientists risk failing to understand how humanity fits into the universe. Our laws that describe nature don't do well at explaining human experience and the modes of thought that explain experience do not constitute "science." However, the very notion of a distinction between between the intellect and the senses, between the laws that govern the universe and the laws that govern human behaviour, is itself an artifact of epistemically singular metaphysics. For example, the Palawa, one of the main groups of Aboriginal Tasmanians, made no distinction between cosmology and code of conduct. The "Palawa used the stars and clouds to determine when to fish, build huts, and travel" and that "the night sky" was "a blackboard on which traditions are drawn with stars, retold to educate generations about moral code and law."[53] This knowledge system was eradicated along with the Palawa people in 19th and 20th centuries.

The dualist schism between the intellect and the senses in modern thought suggests that a non-modern perspective that does not contain a dichotomy between *nomos* and *physis*, between the observer and the observation—might prove helpful in understanding epistemically plural knowledge systems. An example of a living, non-Western scientific system is the one used by the Inuit People who have lived in the Arctic for centuries, if not millennia. The Inuit have migrated across the northern Arctic coast over the last 800 years, having originated on what is now known as the North Slope of Alaska and maintain a measure of genetic continuity with the Paleo-Eskimos who, 4,500 years ago, "spread eastward from Beringia to Greenland, forming small settlements along the Alaskan and Canadian

coasts."[54]

The Inuit people have a concept of law that does not divide the world into natural and non-natural/human or create separation between intellect and senses. Instead, "the Inuit have developed a sophisticated and holistic system of legal mechanisms that" performs "the same function as law in 'Western' cultures"[55] with "juridical functions...entrenched in song-duels, dancing, music and mythological narratives."[56] The Inuit people refer to their knowledge system as Indigenous Knowledge (IK). IK is defined by the Inuit of Alaska, Canada, Greenland, and Chukotka as

> a systematic way of thinking applied to phenomena across biological, physical, cultural and spiritual systems. It includes insights based on evidence acquired through direct and long-term experiences and extensive and multigenerational observations, lessons and skills. It has developed over millennia and is still developing in a living process, including knowledge acquired today and in the future, and it is passed on from generation to generation.[57]

The Inuit apply IK in making food security decisions for the community. Food security involves all aspects of life—including educating the next generation. IK is a scientific system, a model for making predictions about how nature will respond to experiments, that uses individual experiences, instead of measurement, as its fundament. It is epistemically plural. It integrates observers into their observations and, thus, each participant in the IK process makes use of his own unique experience and perspective while recognizing that the other participants also assign unique meanings to their observations. For example, while modern observers consider the night sky at a given place and point in time to be an objective phenomenon, the Inuit do not. The archaeoastronomer Clive Ruggles explains

> Traditional knowledge of the skies can be very localised, and even personal, as among the Inuit, for whom knowledge of celestial phenomena is "in varying degrees, specific to communities, families ... When imparting information, elders frequently made it plain that they were speaking for themselves, that their opinions were not necessarily correct in any absolute sense, and that other elders might, and in probability did, have different views."[58]

The Inuit do not establish a knowledge/belief dichotomy and thus have no need to imagine a "view from nowhere,"[59] to see over the rift. However, without the construct of objectivity, there is no possibility, as there is in an epistemically singular *nomos*, of each researcher having at least theoretical access to the entire corpus of knowledge. One implication, there is no possibility that an epistemically plural knowledge system can outlive the society that created it. In IK, the entire corpus of knowledge exists only in communal form, since each perspective is a unique and complete experience. IK uses the unique perceptions and family teachings of each elder to make complex food security decisions that incorporate predictions about numerous complex phenomena, including weather and the migration patterns of various species. Consistent with IK's observer-integrated approach, IK's predictions are made and tested without sextants, chronometers, or other devices that capture and communicate objective data. Also consistent with IK's observer-integrated approach, the elders' food security predictions—which integrate communal testing plans—are informed by an understanding of nature as a unified system that includes humanity. IK focuses on understanding how subsystems such as air, water, and humanity work in concert and takes into account complex interactions within the biosphere. For example, "changes in sea ice coverage, thickness and timing of formation cause changes in ocean currents, intensity of storms, increase storm surges, distribution of marine flora and fauna, prey dynamics...accessibility to hunting locations, and traveling and hunting safety, all of which require adjustments in hunting and processing strategies."[60] Anthropogenic climate change and government regulations are changing regularities that the region has conformed with for centuries, applying new forms of counterfactual robustness testing to IK.[61]

Scholars of the history and philosophy of science have asked "What would decolonized science look like?"[53] One answer: Inuit Indigenous Knowledge. The survival of the Inuit people attests to the reliability of epistemically plural science for making and testing predictions about nature; the Arctic does not grant mulligans.

**Our Pluralistically Participatory Universe**

IK demonstrates that nature can be understood and navigated through an epistemically plural knowledge system and provides evidence in support of Standish's conjecture "that all of physics may be reducible to properties of the observer." Next, we can marry Standish's conjecture with Holmes's notion that we navigate the world by betting on the chance of not incurring harm. This leads to an observer-centric approach for navigating the world based on harm minimization. We can recast Holmes's prediction system in economic terms and view law as rational choice. Nonetheless, it is difficult to understand how an observer-centric knowledge system could function no matter how rational the choices based on it. The most obvious problem with a subjective knowledge system is that observations may be personal but physical reality is not, we presume some sort of *unus mundus* monism even though we may perceive it through distinct physical and psychological pathways.[62] Even though each Inuit elder might understand the meaning of the night sky in a personal way, nature doesn't care. Different cultures interpret the night sky in different ways but the stars remain the same no matter who observes them. When two physicists, Alice and Bob, look at the same system, let's say a photon, and measure it in the same way, they gain the same "true" knowledge about the system. The same physical reality underlies both observations. Unless Alice and Bob obtain conflicting measurements which would suggest that reality is more complicated than we realize and less objective than we imagine.

Experimental results from Heriot-Watt University demonstrate that Alice and Bob can obtain conflicting measurements about a particle. Massimiliano Proietti and his colleagues tested to see if "two observers can experience seemingly different realities" and found that they can.[63] The researchers explain that "the lack of objectivity revealed by a Bell-Wigner test does not arise in anyone's consciousness, but between the recorded facts"[64] and that the experiment's "measurement records are in conflict regardless of the size or complexity of the observer that records them."[65] The experiment raises the possibility "that different observers irreconcilably disagree about what happened in an experiment"[66] even when "we can tell others what we have done and what we have learned." This is not so difficult to understand when we recognize that there is more to reality than can be accommodated in any single perspective because more than a single perspective is involved in its creation. The Proietti results provide experimental support for Standish's conjecture about physics being reducible to the observer. The results also suggest that Alice and Bob might find it useful to have an understanding of science that allows Alice to understand her own observations but and to make intellectual room for—and sense of—Bob's conflicting observations.

There is an another problem with an observer-centric knowledge system; even the best choices do not always yield the predicted results. Inherent in Holmes's view—that although "we don't know whether anything is necessary or not," we can rationally bet on the consequences of our actions—is a recognition that no decision-making process can eliminate unexpected consequences. Indeterminism—the gap between probability and actuality—is inherent in Holmes's definition of law as a prophesy of how the courts will act, not as the actions of the courts. For Justice Holmes, law exists only as a prediction enforced via expectation. One way of understanding the relation between personal probabilities and physical actualities comes from a 1974 notebook of John Archibald Wheeler. In his notes, Wheeler sketched his thoughts on the centrality of the independent "participator" to physics and advanced the idea that the quantum principle is "participation" and that "physics is not machinery…."[67]

Wheeler prepared the notes on a TWA flight in preparation for a meeting with the logician Dana Scott as well as with Simon Kochen, Roger Penrose, and Charles Patton, a mathematics student with whom Wheeler would jointly present a paper at the First Oxford Conference on Quantum Gravity in a little more than a week and to whom Patton would dedicate his mathematics dissertation three years later.[68] The paper asks a question that is at the core of Wheeler's in-flight thoughts—"is physics legislated by cosmogony?"[69] More than forty years later, Gianfranco Basti described the Patton-Wheeler paper as "visionary" and "in front of the 'quantum information revolution'" spurred by Feynman.[70] Wheeler himself described the most useful outcome of the paper as "a criterion of selection of ideas, exclusion of ideas, based on whether the idea in question admits of having an origin and passing out of existence."[71] Wheeler's thinking at this time was in response to his concern that in the search for "something that might be the foundation for general relativity" that his direction "was too much mathematical, and not enough idea-driven."[72]

Although Wheeler's participatory concept of physics has been described as "*wildly-speculative,*"[73] Walter Elsasser, who received a Ph.D. in quantum mechanics from Max Born and James Franck and served as postdoc research assistant first to Paul Ehrenfest and then Wolfgang Pauli,[74] derived a participator-centric understanding of physics by following Bohr's thinking. Instead of starting with Wheeler's foundational approach to physics, he followed Bohr's "private and not so secret passion" in "the application of atomic physics to biology."[75] This was a path that few physicists followed but for Bohr it was a natural since his father had been a professor of physiology at the University of Copenhagen and he "'grew up in the biological laboratory.'"[76] Elsasser's own interest in quantum biology was spurred by Edington's insistence that the "chief business of the philosopher of science is to eliminate metaphysical arguments from scientific discussions.[77] The problem for Elsasser was that "relativity and quantum theory, have gone far as far as one can in eliminating metaphysics in favor of the use of abstract structures for depicting experience"[78] and the remaining metaphysics presented first Bohr and then Elsasser with an apparently insurmountable difficulty, what Elsasser describes as a "mountain topped with a citadel."

The problem that stumped them is the dualistic schism in how we understand the world, a schism that began in Babylonia and has metamorphosed as its influence spread until it was "codified as philosophical axiom or dogma" by Descartes in the mid-17th century. Elsasser notes that this dualism has gone by many names at many times including "the dualism of body and soul, or of matter and mind"[79] or of ontic and epistemic. His problem with dualist metaphysics, an upset he traces back to Eddington, is that it "is blatantly unscientific" since there are "two qualities in dualism as it is formulated by Descartes, but qualities have no business in science which deals only with relationships. This discrepancy is so deep that it cannot be cured and should not be papered over either."[80]

The discrepancy not only cannot be cured or papered over, it is so deep that it can only be resolved by fixing the underlying problem which is the presumption that an epistemically singular knowledge system can adequately describe physical reality. Singular knowledge systems account for multiple "true" observations of a phenomenon by splitting into separate intellectual and experiential channels which have, at best, an uneasy relation with each other. Dualisms are created by trying to maintain a distinction between the intellect and the senses, *nomos* and *physis*, intelligence and chaos. But the world can be understood and navigated without dualist metaphysics as evidenced by Inuit Indigenous Knowledge and supported by Proietti's findings. A physical universe absent a transcendental verity fits within William James's concept of "a *that* which is not yet any definite *what*, tho' ready to be all sorts of whats; full both of oneness and of manyness, but in respects that don't appear; changing throughout, yet so confusedly that its phases interpenetrate and no points, either of distinction or of identity, can be caught.[81]

It is easy to criticize dualist metaphysics but is it feasible to get along without them? Elsasser explained at a 1978 MIT entropy conference that "only an utter fool could make a frontal attack on a mountain topped by a citadel."[82] Not being a fool, Elsasser followed the "normal strategy" which "is to circumvent the obstacle."[83] His route "was to find concepts specific for biology but do not resist being treated as operational or otherwise formal."[84.] Elsasser explains that the biological

> concept that I discovered is that of individuality. It has been grossly neglected by scientists and philosophers alike. It has no place in the world of the physicist but there is no biologist who, if you listen to him long enough, will not tell you that no two organisms are ever exactly alike. I use the concept in such a way that it should better be put into the plural, individualities. I think of an organism as a collection of individualities at all levels….
>
> Bohr has always been searching for a generalization of the physicist's complementarity. The notion of individuality give us such a generalization. The homogeneity of ordinary physical objects -- is mutually exclusive with the collection of individualities that form a living thing.[85]

The link between individual biological systems and quantum theory was formalized in a 2002 paper by Carlton M. Caves, Christopher A. Fuchs, and Rüdiger Schack.[86] The paper demonstrated that quantum probabilities are Bayesian probabilities; the subjective predictions produced by a biologic system are quantum probabilities. The paper provides the theoretical basis for participatory physics and supports

Elsasser's conclusions. Biocentric understandings of quantum theory provide a physical explanation for experiential reality. These understandings are epistemically plural since they recognize that individuals have unique perspectives. Fuchs explicates a biocentric understanding of quantum theory by arguing that every "time an act of observer-participancy occurs (every time a quantum measurement occurs)...the life of the universe as a whole takes on a deeply new character…."[87] If we substitute observation or experience for the term quantum measurement, Fuchs's argument can be thought of as a dynamic version of Holmes's definition of law: every time an individual makes an observation, law is updated. Wheeler sums up his view with the statement "No participator, no world!"[68] This can be restated as "no participator, no law" which is consistent with Holmes's definition of law as the participator's prophesy. The updating of law means that there is in an ongoing flow of revised possibilities for the participator to choose from. Reality is fertile. When we interact with it we create new reality. Fuchs uses the term "participatory realism" to describe how "the world is so wired that our actions as active agents actually matter. Our actions and their consequences are not eliminable epiphenomena."[88]

Participatory realism is an affirmative answer to Wheeler's speculative suspicion about the world we live in—"A participatory universe?"[89] Fuchs is an expounder of an understanding of quantum mechanics, QBism,[90] that makes a foundational statement about reality, including the role of you, my reader, in creating it right now. QBism breaks with objectivation and, with it, the big-bang-rules-all cosmogony. Instead, QBism emphasizes the interactions of a multitude of individual, autonomous agents in an ongoing process of creation. The QBist world is epistemically plural. The term "Qblooey – *cf. kablooey*" is used by QBists as "a slur on the usual conception of the Big Bang, where the universe had its one and only creative moment at the very beginning."[91] By breaking with objectivation, QBism offers the first decolonized explication of physics. The physicist Hans Christian von Baeyer resolves

Democritus's conundrum as he summarizes the QBist approach to quantum mechanics by describing it as the Intellect beginning to respect the Senses.[92]

The quantum nature of the world is often thought to apply at only the tiniest of scales. However, Fuchs and the mathematician Rüdiger Schack explain that "quantum mechanics can be applied to any physical system...all the way to living beings and other agents."[93] Because reality is in an ongoing creative process, "law" can be only predictive, not determinative. This is the genesis of indeterminism.

The idea of scientific law as non-determinative can be seen in Schack's statement that there "are no laws of nature…."[94] Schack's statement reflects an alternative understanding of Einstein's 1942 letter to Fritz and Berta Reiche in which he wrote "I still do not believe that the Lord God plays dice. If he had wanted to do this, then he would have done it quite thoroughly and not stopped with a plan for his gambling: In for a penny, in for a pound (*Wenn schon*, *denn schon*). Then we wouldn't have to search for laws at all."[95] Schack argues there "are indeed no laws…. There are no laws of nature, not even stochastic ones. The world does not evolve according to a mechanism." He explains that, instead of determinative laws, what we have are "tools for agents to navigate the world, to survive in the world."[96] Quantum mechanics is a formalization of these tools "for ordering and surveying human experience." Carlo Rovelli explains that although science "is extremely reliable; it's not certain. In fact, not only is it not certain, but it's the lack of certainty that grounds it."[97] Holmes concludes that "it often is a merit of an ideal to be unattainable. Its being so keeps forever before us something more to be done, and saves us from the ennui of a monotonous perfection."[98]

**Conclusion**

We looked at whether a scientific system can function without objective data and found that it can. We looked at the consequences of objectivation and found that it means we know less about ourselves then we know about the natural world which means that we know less about the natural world than we think. Herein lies the alpha of the gaps, lacunae, paradoxes and antinomies that expose objectivation's limitations. Which leaves us where? How do individuals participating in an epistemically singular *nomos* navigate a quantum universe? Cover notes that the "challenge presented by the absence of a single, 'objective' interpretation is, instead, the need to maintain a sense of...meaning despite the destruction of any pretense of superiority of one *nomos* over another."[99] He concludes "We ought to stop circumscribing the *nomos*; we ought to invite new worlds."[100]

What are we talking about when we talk about quantum mechanics? We are talking about a knowledge system that dismisses Cartesian dualism by embodying a pluralistic concept of truth. In doing so, it poses the most profound challenge to the dominant scientific paradigm since Christoph Scheiner and Galileo Galilei discovered sunspots in 1611 and proceeded to debate the physical meaning of their observations.[101] Which is why people care about what we talk about when we talk about quantum mechanics.

**References**

1 Walter M. Elsasser, “Induction and the Two Cultures”, in Raphael D. Levine and Myron Tribus, eds., The Maximum Entropy Formalism (Cambridge and London: The MIT Press, pp. 211-217, p. 216.

2 A slight recasting of N. David Mermin’s philosophical question, “What the hell are we talking about when we use quantum mechanics?” in N. David Mermin, “Making better sense of quantum mechanics”, arXiv:1809.01639 [quant-ph], 5 September 2018, p. 3.

3 Adán Cabello, “Interpretations of quantum theory: A map of madness,” arXiv:1509.04711v2 [quant-ph], 20 November 2016, Abstract.

4 For a informal (out of date) short-list of schools of quantum interpretation as well as a list of efforts to achieve consensus, see Christopher A. Fuchs, “Quantum Mechanics as Quantum Information (and only a little more), arXiv:quant-ph/0205039 [quant-ph], 8 May 2002, pp. 1-2. For a more formal classification of quantum interpretations, see Cabello, p. 2.

5 See Feynman’ statement “And I'm not happy with all the analyses that go with just the classical theory, because nature isn't classical, dammit, and if you want to make a simulation of nature, you'd better make it quantum mechanical, and by golly it's a wonderful problem, because it doesn't look so easy” in Richard P. Feynman, “Simulating Physics with Computers”, *International Journal of Theoretical Physics* 21:6/7 (1982), pp. 467–488 on 486.

6 Plato, *Gorgias* 454d, Perseus Digital Library, Tufts University, http://www.perseus.tufts.edu/hopper/text?doc=Perseus%3Atext%3A1999.01.0178%3Atext%3DGorg.%3Asection%3D454d

7 Niels Bohr, “On the Application of the Quantum Theory to Atomic Structure: Part I. The Fundamental Postulates”, *Proceedings of the Cambridge Philosophical Society: Supplement*, (Cambridge: At the University Press, 1924), p. 1.

8 Niels Bohr, “On the Notions of Causality and Complementarity”, *Science*, New Series, Vol. 111, No. 2873 (Jan. 20, 1950), pp. 51-54 on 52.

9 Ibid.

10 Bohr, “On the Application of the Quantum Theory to Atomic Structure,” (ref. 7), p. 1.

11 Gabriele Lolli, The Meaning of Proofs, Mathematics as Storytelling, (trans.) Bonnie McClellan-Broussard (Cambridge and London: The MIT Press), p. xiv.

12 Ray P Norris and Barnaby R.M. Norris, “Why Are There Seven Sisters?” in Efrosyni Boutsikas, Stephen C. McCluskey, John Steele (eds) Advancing Cultural Astronomy: Studies In Honour of Clive Ruggles, (Springer, Historical & Cultural Astronomy, 2021) pp. 223-235 on 234.

13 See Patrick D. Nunn. “Australian Aboriginal Traditions about Coastal Change Reconciled with Postglacial Sea-Level History: A First Synthesis.” *Environment and History* 22:3 (2016), pp. 393-420 on pp. 417-18.

14 Patrick D. Nunn and Nicholas J. Reid, “Aboriginal Memories of Inundation of the Australian Coast Dating from More than 7000 Years Ago,” *Australian Geographer* 47:1 (2016) pp. 11-47 on 13.

15 Ibid, p. 419 (note omitted).

16 Bohr, “On the Notions of Causality and Complementarity” (ref. 7), p. 53.

17 Graham Harman, “The Importance of Bruno Latour for Philosophy”, *Cultural Studies Review,* 13:1 (2007), 31-49 on 46.

18 Susan Sontag, Illness as Metaphor (New York: Picadore, 1978), chap. 7, footnote 2.

19 See Russell K. Standish, “The importance of the observer in science,” In Proceedings *The Two Cultures: Reconsidering the division between the Sciences and Humanities*, University of New South Wales (2005), arXiv:physics/0508123 [physics.pop-ph], Abstract.

20 Ibid.

21 Hans Christian von Baeyer, QBism, The Future of Physics (Cambridge: Harvard University Press, 2016), p. 186 (noted omitted).

22 Maria Carla Galavotti, “Subjectivism, Objectivism and Objectivity in Bruno de Finetti’s Bayesianism,” In *Foundations of Bayesianism*, Vol 24 of the Applied Logic Series, David Corfield and Jon Williamson, ed., 161-174 (Dordrecht: Springer Science+Business Media, 2001), p. 169).

23 Ibid.

24 Ibid., p. 171.

25 Suzi Adams, “The Enduring Enigma: *Physis* and *Nomos* in Castoriadis”, *Thesis Eleven* 65:1 (2001), 93-107 on 95.

26 Henry Goodwin Smith, “Persian Dualism”, *The American Journal of Theology*, 8:3 (1904), pp. 487-501 on 490.

27 Erwin Schrödinger, ‘Nature and the Greeks’ and ‘Science and Humanism’ (Canto original series). (Cambridge: Cambridge University Press, 2014), p. 89.

28 Johnjoe McFadden and Jim Al-Khalili, “The origins of quantum biology”, *Proceedings of the Royal Society A*, 474:2220 (Dec. 2018), p.11.

29 Schrödinger, ‘Nature and the Greeks’, (ref. 24), p. 92.

30 Ibid.

31 Benjamin Madely, “Patterns of frontier genocide 1803–1910: the Aboriginal Tasmanians, the Yuki of California, and the Herero of Namibia”, *Journal of Genocide Research* 6:2 (2016), pp. 167–192 on 169.

32 Ibid. (note omitted).

33 Ibid., p. 167.

34 Valentin Jeutner, “Laws of Nature and Nature of Law”, The Legal Dimensions of Quantum Computing Conference: Panel 5 – Quantum legal theory (Lund University), April 29, 2022.

35 Robert M. Cover, “Forward: Nomos and Narrative”, *Harvard Law Review* 97:4 (1983-1984), pp. 4-68 on 42.

36 Ibid., p. 68.

37 Oliver Wendell Holmes, “The path of the law”, *Harvard Law Review*, 10:8 (1897), 457-78, on 460.

38 Ibid., p. 461.

39 Christopher A. Fuchs and Blake C. Stacey. “QBism: Quantum Theory as a Hero’s Handbook”, Proceedings of the International School of Physics “Enrico Fermi” Vol. 197, Foundations of Quantum Theory (2016), p. 1.

40 Michael F. Duggan, “The Municipal Ideal and the Unknown End: A Resolution of Oliver Wendell Holmes” *North Dakota Law Review* 83:2 (2007), 463-546, on 515.

41 Richard A, Posner, ed., Essential Holmes: Selections from the Letters, Speeches, Judicial Opinions, and Other Writings of Oliver Wendell Holmes, Jr. (Chicago: The University of Chicago Press, 1996), p. 108.

42 Oliver Wendell Holmes, “Law In Science and Science in Law,” *Harvard Law Review*, 12:7 (1899), 443-463, on 447.

43 Dean Buonomano and Carlo Rovelli, “Bridging the neuroscience and physics of time”, arXiv:2110.01976 (2021), p. 1.

44 Ibid.

45 Ibid.

46 Ibid., p. 7.

47 Bruno Schultz quoted in J. M. Coetzee, Inner Workings: Literary Essays, 2000-2005. 1st American ed. (New York: Viking, 2007), pp. 72-73.

48 Kathrin Busch, “Artistic Research and the Poetics of Knowledge”, *Art and Research: A Journal of Ideas, Contexts and Methods*, 2:2 (2009), pp. 1-7, on 3.

49 Ibid., p. 4.

50 Ibid.

51 Erwin Schrödinger, “Discussion of Probability Relations between Separated Systems”, *Mathematical Proceedings of the Cambridge Philosophical Society*, 31:4 (1935), pp. 555-563 on 555.

52 Ibid.

53 Michelle Gantevoort, Duane W. Hamacher, and Savannah Lischick, “Reconstructing the Star Knowledge of Aboriginal Tasmanians,” *Journal of Astronomical History and Heritage*, 19:3, Preprint (2016), pp. 13, 26.

54 Jennifer A. Raff, Margarita Rzhetskaya, Justin Tackney, and M. Geoffrey Hayes, “Mitochondrial diversity of Iñupiat people from the Alaskan North Slope provides evidence for the origins of the Paleo- and Neo-Eskimo peoples,” *American Journal of Physical Anthropology,* 157:4 (2015), pp. 603–614 on 603.

55 Natalia Loukacheva, “Indigenous Inuit Law, ‘Western’ Law and Northern Issues.” *Arctic Review on Law and Politics* 3:2 (2012), pp. 200–217 on 202.

56 Ibid., p. 205.

57 Inuit of Alaska, Canada, Greenland, and Chukotka, “Utqiaġvik Declaration,” July 16-19, 2018, Utqiaġvik, Alaska, pp. 6-7.

58 Clive L.N. Ruggles, “Indigenous Astronomies and Progress in Modern Astronomy,” in Ray P. Norris and Clive L.N. Ruggles, eds., Accelerating the Rate of Astronomical Discovery - sps5, Proceedings of Science-Special Session 5, IAU General Assembly, (Rio de Janeiro, Brazil, August 11-14, 2009), p. 9.

59 Thomas Nagel, The View From Nowhere, (New York, Oxford: Oxford University Press, 1986).

60 Inuit Circumpolar Council-Alaska, “Alaskan Inuit Food Security Conceptual Framework: How to Assess the Arctic From an Inuit Perspective. Technical Report,” (Anchorage: Inuit Circumpolar Council-Alaska, 2015), p. 66.

61 Ibid., pp. 65-66.

62 Harald Atmanspacher provides a formal outline of dual-aspect monism (mind-body dualism) based on the Pauli–Jung conjecture in Harald Atmanspacher, “The Pauli–Jung Conjecture and Its Relatives: A Formally Augmented Outline”, Open Philosophy, 3:1 (2020), pp. 527-549 on 529-532.

63 Massimiliano Proietti, Alexander Pickston, Francesco Graffitti, Peter Barrow, Dmytro Kundys, Cyril Branciard, Martin Ringbauer, and Alessandro Fedrizzi, "Experimental test of local observer independence," *Science Advances* 5:9 (2019), pp.1-6 on 1.

64 Ibid., p. 4.

65 Ibid.

66 Ibid.

67 John A. Wheeler, "Add 'Participant' to 'Undecidable Propositions' to Arrive at Physics," In-flight notes, draft (4-6 February 1974), on p. 1. Retrieved from https://jawarchive.files.wordpress.com/2012/03/twa-1974.pdf.

68 Charles Patton, "Classification of Relativistic n-Particle Dynamics", Dissertation for the degree of Doctor of Philosophy in Mathematics, State University of New York at Stony Brook (1977), p. v.

69 Charles M. Patton and John A. Wheeler, "Is physics legislated by cosmogony?" In Christopher J. Isham, Roger Penrose, and Dennis W. Sciama (eds.), *Quantum gravity; Proceedings of the Oxford Symposium, Harwell, Berks., England, February 15, 16, 1974* (Oxford: Clarendon Press, 1975), p. 538.

70 Gianfranco Basti, "From formal logic to formal ontology: The new dual paradigm in natural science," in Fábio Bertato, ed. *Proceedings of the 1st CLE Colleoquem for Philosophy and History of Formal Sciences, Campinas, 21-23 March 2013 (Campinas, 2014)* in press, Abstract.

71 John Wheeler, Oral History, Session XII, Interviewed by Kenneth W. Ford at Jadwin Hall, Princeton University, 1994 March 28, retrieved from www.aip.org/history-programs/niels-bohr-library/oral-histories/5908-12.

72 Ibid.

73 Christopher A. Fuchs, "On Participatory Realism," arXiv:1601.04360v3 [quant-ph] (2016), footnote 7.

74 See Bruce D. Marsh, “Memorial to Walter M. Elsasser: 1904-1991, *Geological Society of America Memorials*, v. 24, pp. 181-186, p. 181.

75 Elsasser, “Induction and the Two Cultures”, (ref. 1), p. 215.

76 Ibid., p. 216.

77 Ibid., p. 215.

78 Ibid.p.

79 Ibid., p. 216.

80 Ibid., pp. 216-217.

81 William James, “The Thing and Its Relations” *The Journal of Philosophy, Psychology and Scientific Methods,* II:2 (1905), pp. 29-41 on 29.

82 Elsasser, “Induction and the Two Cultures”, (ref. 1), p. 217.

83 Ibid.

84 Ibid.

85 Ibid., p. 217.

86 Carlton M. Caves, Christopher A. Fuchs, and Rüdiger Schack, “Quantum probabilities as Bayesian probabilities”, *Phys. Rev. A* 65, 022305, (2002).

87 Fuchs, “On Participatory Realism,” (ref. 72), footnote 7.

88 Fuchs, “On Participatory Realism” (ref. 72), p. 11.

89 Von Baeyer, QBism (ref 19), p. 204.

90 née Quantum Bayesianism. However, there are crucial distinctions between QBism and its intellectual progenitor, see Blake C. Stacey, “Ideas Abandoned en Route to QBism.” (2019) arXiv:1911.07386v2.

91 Fuchs, “On Participatory Realism”, (ref. 72), p. 10.

92 Von Baeyer, QBism (ref 19), p. 155.

93 Christopher A. Fuchs and Rüdiger Schack, "QBism and the Greeks: why a quantum state does not represent an element of physical reality", arXiv:1412.4211v2 [quant-ph] (2015), p. 4.

94 Fuchs, "On Participatory Realism" (ref. 72), footnote 6.

95 Ibid.

96 Ibid.

97 Carlo Rovelli, "Science Is Not About Certainty", *New Republic*, (July 11, 2014).

98 Oliver Wendell Holmes, "Law In Science", (ref. 40), 463.

99 Cover, "Nomos and Narrative", (ref. 33), p. 44.

100 Ibid., 68.

101 See David Marshall Miller's review of *On sunspots: Galileo Galilei and Christoph Scheiner*, translated and introduced by Eileen Reeves and Albert Van Helden, Aestimatio (2012), pp. 97-102 on 97.